\documentclass[apj,twocolappendix, numberedappendix]{emulateapj}

\usepackage{textcomp}
\usepackage{amsmath}
\usepackage{graphicx}
\usepackage{subfigure}
\usepackage{color}
\usepackage{verbatim}
\usepackage{bm}% bold math
\usepackage{hyperref}

\hypersetup{colorlinks = true, linkcolor = blue, anchorcolor = red, citecolor = blue, filecolor = red, pagecolor = red, urlcolor = red}

\begin{document}

%\title{A statistical study of X-ray flares in Sgr A$^\star$: solar-like self-organization criticality }
%\maketitle

\title{Exploring the accretion model of M87 and 3C 84 with the Faraday rotation measure observations}

\author
{Ya-Ping Li$^{1}$, Feng Yuan$^{2,1}$, Fu-Guo Xie$^2$}
\affil{$^{1}$Department of Astronomy and Institute of Theoretical Physics and Astrophysics,
Xiamen University, \\ Xiamen, Fujian 361005, China;}
\affil{$^{2}$Key Laboratory for Research in Galaxies and Cosmology, Shanghai Astronomical Observatory, Chinese Academy of Sciences, 80
Nandan Road, Shanghai 200030, China; fyuan@shao.ac.cn (FY)}

\begin{abstract}
Low-luminosity active galactic nuclei (LLAGNs) are believed to be powered by an accretion-jet model, consisting of an inner advection-dominated accretion flow (ADAF), an outer truncated standard thin disk, and a jet. But model degeneracy still exists in this framework. For example, the X-ray emission can originate from either the ADAF or jet. The aim of the present work is to check these models with the Faraday rotation measure (RM) observations recently detected for two LLAGNs, M87 and 3C 84, in the sub-mm band. For M87, we find that the RM predicted by the model in which the X-ray emission originates from the ADAF is larger than the observed upper limit of RM by over two orders of magnitude, while the model in which the X-ray emission originates from the jet predicts a RM lower than the observed upper limit. For 3C 84, the sub-mm emission is found to be dominated by the jet component, while the Faraday screen is attributed to the ADAFs. This scenario can naturally explain the observed {\it external} origin of the RM and why RM is found to be stable during a two-year interval although the sub-mm emission increases at the same period.
\end{abstract}

\section{introduction}

Some of the most important information about the physics of AGNs comes from their spectral energy distribution (SED). For the most nearby low-luminosity AGNs (LLAGNs),  in addition to the extremely low Eddington ratios ($L_{\rm bol}/L_{\rm Edd}\lesssim10^{-2}$, where $L_{\rm bol}$ and $L_{\rm Edd}$ are, respectively, the bolometric and Eddington luminosity), there are some remarkable features in comparison with luminous AGNs, e.g., the absence of the big blue bump, the weakness or the lack of reflection features in the X-ray band, a relative large X-ray-to-optical ratio ($\alpha_{\rm ox}\gtrsim-1.0$), the narrowness of the iron K$\alpha$ line, the frequent detection of double-peaked broad Balmer lines, and the prevalence of low-ionization nebular conditions (e.g., \citealt{Ho99,Ho09,Ho00,Terashima02,Ptak04}, see \citealt{Ho08} for a review). All these features point to a truncated disk scenario, namely an outer optically thick, geometrically thin disk \citep{SS73} is truncated at a certain radius and is replaced by an inner hot accretion flow or advection-dominated accretion flow (ADAF; \citealt{Narayan94}; see review by \citealt{Yuan14}).
Another intriguing feature of LLAGNs is the ubiquity of compact radio cores or jets, whose radio strength often qualifies them as being ``radio loud" \citep{Falcke00,Ho02}. The radio emission is generally far greater than can be attributed to an ADAF and is more consistent with a jet origin.  Therefore, a three-components model, consisting of an inner ADAF, an outer ``truncated" standard thin disk, and a radio jet (hereafter coupled accretion-jet model or ADAF-jet model), is preferred for modeling the SEDs of LLAGNs \citep{Yuan14}.

Although the general success of the coupled accretion-jet model \citep{Yuan14}, some model degeneracy still exists, and the modeling to the SEDs is not unique. The origin of X-ray emission is different in different models. Taking a nearby LLAGN -- M87--- as an example, in the ``jet-dominated'' model, the X-ray emission is dominated by the jet \citep{Yuan09,Yu11,Nemmen14}, while the X-ray spectrum can also be well explained by the ADAF in the ``ADAF-dominated'' model \citep{DiMatteo03,Nemmen14}. The physical argument underlying the ``jet-dominated'' model was stated in \citet{YC05}. In general, both the ADAF and the jet contribute to the X-ray emission. The X-ray emission from the ADAF is roughly proportional to $\dot{M}^2$ while that from the jet to $\dot{M}_{\rm jet}$. Here $\dot{M}$ and $\dot{M}_{\rm jet}$ are the mass accretion rate of the ADAF and the mass lost rate in the jet, respectively. If $\dot{M}$ and $\dot{M}_{\rm jet}$ are proportional to each other, which is reasonable, we expect that with the decrease of $\dot{M}$ (and therefore luminosity), the contribution from the jet to the X-ray emission will become more and more important. Detailed calculations show that when the 2$-$10 keV X-ray luminosity is lower than a critical value $L_{\rm X,crit}$ given by ${\rm log} (L_{\rm X,crit}/L_{\rm Edd})\approx -5.36-0.17{\rm log}(M/M_{\sun})$, the X-ray emission of the system should be dominated by the jet rather than the ADAF \citep{YC05}.
%, which will lead to a steeper radio-X-ray correlation with correlation slop of $\sim1.23$ compared with the canonical index of $\sim0.7$. The critical value is $L_{\rm X,crit}=1.0\times10^{-7}\ L_{\rm Edd}$ for M87, while its observed X-ray luminosity $L_{\rm X}=2.3\times10^{-8}\ L_{\rm Edd}$.
The prediction is supported by later observations (e.g., \citealt{Pellegrini07,Wu07,Wrobel08,Yuan09,deGasperin11,Younes12}). The X-ray luminosity of M87 is below this critical value and this is the physical base of the jet-dominated model. In fact, based on the similar X-ray spectrum of the nucleus with the jet knots, \citet{Wilson02} have suggested that the X-ray emission in M87 may originate from the (sub)pc-scale jet rather than the ADAF.

The measurement of rotation measure (RM) supplies an additional constrain to the model thus it can potentially test the existing models and break the model degeneracy. Faraday rotations have been detected for a number of pc-scale AGN jets at several low frequency radio bands (e.g., \citealt{Owen90,Zavala03,Hovatta12,Algaba16,Pasetto16}).
%probe the innermost region of the AGNs.
%With the recourse of the polarization data, we sometimes can break the degeneracy of the accretion model.
These observations are not useful to discriminate the above-mentioned two models since in both models the radio emission comes from the jet. However, it has been very difficult to obtain accurate polarimetric information at sub-mm wavelength due to the instrumentation limit. Previous efforts to detect RM at sub-mm wavelengths from AGNs only result in several successful detections. One is the detection of RM in our Galactic center SMBH, Sgr A$^\star$, which is $(-5.6\pm0.7)\times10^{5}\ {\rm rad~m}^{-2}$ from the simultaneous observations at multiple frequencies using the Submillimeter Array polarimeter \citep{Marrone07}. Two models of Sgr A$^\star$ are the hot accretion flow model \citep{Yuan03} and the jet model \citep{FM00}. \citet{Li15} have calculated the predicted RM of these two models and compare with the observed value. It is found that the predicated RM from the jet model is found to be two orders of magnitude lower than the measured value. The RM observations thus put strong challenge to the jet model of Sgr A$^\star$.

Recently, measurements of Faraday rotation at $0.9-1.3$ mm have been reported for two nearby AGNs\footnote{\citet{Marti-Vidal15} detect high RM for a distant gravitationally lensed AGN, PKS $1830-211$, using the Atacama Large Millimeter/submillimeter Array (ALMA). Due to uncertainties for the broad-band spectra, we defer the study for PKS 1830-211 in the future.}. One is 3C 84,  with the ${\rm RM}=(8.7\pm6.9)\times10^{5}\ {\rm rad~m}^{-2}$ (3$\sigma$ uncertainty; \citealt{Plambeck14}). Another one is M87, with ${\rm RM}=(-2.1\pm5.4)\times10^{5}\ {\rm rad~m}^{-2}$ (3$\sigma$ uncertainty; \citealt{Kuo14}). Note that the RM detected for M87 can only be regarded as an upper limit since the $3\sigma$ range crosses zero (\citealt{Kuo14}; private communication with Keiichi Asada). The goal of the paper is to check the accretion-jet model of M87 and 3C 84 using their RM measurements.

The paper is organized as follows. In Section 2, we model the SED of M87 and 3C84. We then calculate the RM and discuss the physical implications for the accretion-jet model in Section 3. We briefly summarize our results in Section 4.

\section{SED modeling}

\subsection{The Accretion-Jet Model}
We only briefly describe the coupled accretion-jet model here, the details of which can be found in \citet{YCN05,Yuan09}. The ``truncation" radius $R_{\rm tr}$ of the accretion flow
%consists of an inner ADAF within a ``truncation" radius $R_{\rm tr}$ and a truncated cool thin disk outside of $R_{\rm tr}$.  The value of $R_{\rm tr}$
mainly determines the emitted spectrum from the outer thin disk, while its effect on the ADAF is very small since the radiation of ADAF comes from the innermost region of the accretion flow.

One of the most important progresses in black hole hot accretion flows in recent years is the theoretical identification of strong wind from accretion flow \citep{Yuan12b,Yuan15,Narayan12,Li13}. The presence of wind results in the inward decrease of mass accretion rate, $\dot{M}(R)=\dot{M}_{\rm out}(R/R_{\rm tr})^s$. The ``wind parameter'' $s$ have been calculated in a number of numerical simulations (e.g., \citealt{Stone99,Hawley02,Igumenshchev03}; see \citealt{Yuan12a} for a review). These simulations show that the value of $s$ is in the range $0.3-1.0$. There are also some analytical studies on wind from accretion flows (e.g., \citealt{Blandford99,Jiao11,Begelman12,Gu15}). \citet{Mosallanezhad16} have recently extended these studies by including magnetic field and consistency has been found between this analytical work and the most recent GRMHD numerical simulation study of \citet{Yuan15}. The existence of wind has been confirmed by the 3M seconds \emph{Chandra} observations of Sgr~A$^\star$ ($s\simeq1.0$; \citealt{Wang13}) and 1M second \emph{Chandra} observation of NGC 3115 ($s\simeq0.5$; \citealt{Wong14}).

In addition to the wind parameter $s$, other parameters of the hot accretion flow include the viscosity parameter $\alpha$, magnetic parameter $\beta$  (defined as the ratio of the gas pressure $p_{\rm gas}$ and the magnetic pressure $p_{\rm mag}$, $\beta\equiv~p_{\rm gas}/p_{\rm mag}$), and the fraction of the turbulent dissipation that directly heats the electrons $\delta$. We adopt the typical values of these parameters widely used in the literatures: $\alpha=0.3$, $\beta=10$, and $\delta=0.5$ unless otherwise indicated.
%For the truncation radius , it is set to be $R_{\rm tr}=10^{4}\ R_{\rm S}$, where $R_{\rm S}\equiv2GM_{\bullet}/c^2$ is the Schwarzschild radius of the black hole.

The one-dimensional dynamics of the ADAF is governed by conservations of mass, radial momentum, angular momentum, electron, and ion energy equations (refer to Equations 1-5 in \citealt{Yuan00,Yuan03} for details). For a given the mass accretion $\dot{M}_{\rm out}$ at $R_{\rm tr}$, we adjust an eigenvalue of the problem, which is the specific angular momentum of the flow at the horizon $j_{0}$, together with the outer boundary conditions at $R_{\rm tr}$, which includes the electron temperature $T_{\rm out,e}$, the ion temperature $T_{\rm out,i}$ and the radial velocity, and the sonic point condition, we obtain the global solutions of the hot accretion flow. This solution contains all the dynamical properties of the accretion flow, such as the electron density, electron temperature, magnetic field, and the radial velocity.
The radiation from ADAF can then be calculated by taking account into the synchrotron, bremsstrahlung of the thermal electrons in the accretion flow and their Comptonization \citep{Yuan00}.  We follow the method of \citet{Coppi90} to calculate the Comptonization spectrum. For radiative transfer calculation, in which the self-absorption of synchrotron emission is included, the ``plane-parallel rays" method is adopted (e.g., \citealt{Yuan00}).

The radiation from the thin disk is simply a multi-temperature blackbody spectrum with the temperature profile $T\propto R^{-3/4}$, which extends from $R_{\rm tr}$ to $10^5\ R_{\rm S}$. The emission from the outer thin disk is insensitive to the outer boundary of the disk.
%For the sources considered in the work, we found that the contribution from the thin disk to the total radiation can be negligible as shown in our following SED modeling results.   %This large truncation radius suggests a negligible contribution of the outer thin disk to the observed SED, which is also verified by our SED fitting results below.

The jet model is adopted from the internal shock scenario (see the appendix of \citealt{YCN05} for details). Compared with the accretion flow model, there are more uncertainties in the jet model. The main parameters for the jet include the mass loss rate $\dot{M}_{\rm jet}$, the half-opening angle $\phi$, the bulk Lorentz factor $\Gamma$ of the jet, the power-law index $p$ of the accelerated relativistic electrons in the jet, the fraction of electrons accelerated into the power-law distribution by the shock $\xi_{\rm e}$, and the inclination angle $\theta$ of the jet relative to our LOS. We also define in the comoving frame two dimensionless parameter, $\epsilon_{\rm e}$ and $\epsilon_{\rm B}$, to measure the ratio of the energy density of power-law electrons and magnetic fields, respectively, to the shock energy density.  We adopt the ``typical" value of $\phi=0.1$ rad and $\xi_{\rm e}=0.01$. Some parameters, such as $\theta$, $\Gamma$, can be constrained by other independent observations, so they are not free. The emission from the jet includes the synchrotron radiation from both the power-law and thermal electrons after considering the self-absorption effect, although we find that the former plays a dominant role. The total emission is obtain by integrating emission from different locations in the jet.

We then sum the three components to obtain the total radiation and compare it with the observed multi-wavelength spectrum. We adjust the model parameters to make the model spectrum fitting satisfactorily. The three components in the model are coupled with each other so some parameters are not completely free. For example, the mass accretion rate of ADAF at the truncation radius $R_{\rm tr}$ should be the same as the accretion of the outer standard thin disk.  The mass loss rate $\dot{M}_{\rm jet}$ in the jet should also be a reasonably small fraction of  the accretion rate in the innermost region of the accretion flow \citep{Yuan14}.  As the methodology of most of previous works, we judge the goodness of the fit by eyes instead of using any rigorous statistical analysis. This is partly due to the big difficulty in obtaining a transonic global solution of an ADAF. More importantly, we think the simplifications of the one-dimensional theoretical models make such elaborate statistical analysis meaningless.
%make the meaning of  such statistical analysis unclear.

\subsection{SED Modeling of M87}\label{sec:sed:87}

M87 hosts an SMBH with its mass of $M_{\bullet}=6.4\times10^{9}~M_{\odot}$ \citep{Gebhardt09} with a large scale collimated jet. The distance to M87  is assumed to be 16.5 Mpc \citep{Jordan05} in this paper. For the distance and mass of M87, $1''$ in the sky extends to 80 pc $\simeq1.1\times10^5\ R_{\rm S}$.
The broad-band spectral data are mainly collected from \citet{Maoz07,Yu11,Prieto16}.
There are two high resolution radio data sets, one with $0.15''$ resolution, the other with $0.4''$ resolution \citep{Prieto16}. We in this work adopt the $0.15''$ data set, in order to avoid the possible contamination from its host galaxy.
%We  present high angular resolution SEDs of the core of M87 at a scale of both 0.4 arcsec and 0.15 arcsec compiled in \citet{Prieto16} to compare different data sets with each other. We can see that the 0.4 arcsec angular resolution SED data are higher than the other two data sets by a factor of a few in the low frequency radio band. At sub-mm, optical-UV and X-ray bands, they are well consistent with each other. We adopt the SED data with the higher angular resolution ($0.15''$) to compare with our model in order to avoid the possible contamination from the host galaxy.
The small Eddington ratio of the bolometric luminosity, $3.6\times10^{-6}$, together the large scale radio jet observed, justifies the utilization of ADAF-jet model to interpret the data.

Before fitting the observed SED, there are several observational constraints which we should consider. The Bondi accretion rate is estimated to be $\dot{M}_{\rm Bondi}=0.01\sim0.2~M_{\odot}\ {\rm yr^{-1}}$ (scaled to the Eddington accretion rate, $\dot{m}_{\rm Bondi}=6.5\times10^{-5}\sim1.4\times10^{-3}$; \citealt{Churazov02,DiMatteo03,Russell15}). \citet{Biretta99} analysed the superluminal motion of the jet in M87 with \emph{HST} observations and estimated that $\Gamma\geq6$ and $10^{\circ}<\theta<19^{\circ}$ \footnote{\citet{Wang09} estimated the inclination angle for the jet region to be $\sim14^{\circ}$ based on multi-band fitting to radio-to-X-ray continua. Another estimate of $\theta\approx15^{\circ}-25^{\circ}$ was provided by \citet{Acciari09}. These  variations of the angle only lead to a marginal effect in the spectra produced by the jet component.}. Based on these results, we adopt $\Gamma=6$ and $\theta=10^{\circ}$ in our SED modeling. For the truncation radius we set $R_{\rm tr}=R_{\rm Bondi}$, which means that the ADAF can extend all the way up to the Bondi radius. This is because the accretion rate in M87 is very low \citep{Yuan14}. \citet{Russell15} found that the density profile around the Bondi radius at $0.12-0.22$ kpc ($\sim(1.7-3.0)\times10^5~{R_{\rm S}}$) is consistent with $\rho\propto r^{-b}$ with $b\simeq{1\sim1.2}$. This density profile indicates an outflow parameter $s\simeq0.3\sim0.5$. This value of $s$ is somewhat smaller than that obtained by numerical simulations \citep{Yuan12a}. The reason may be because of the low angular momentum of the accretion flow in M87 \citep{Bu13}. In the present work, we adopt two values of $s$, $s=0.3$ in Model A1 and $s=0.5$ in Model A2.

\begin{figure}[htb]
\begin{center}
\includegraphics[width=0.45\textwidth]{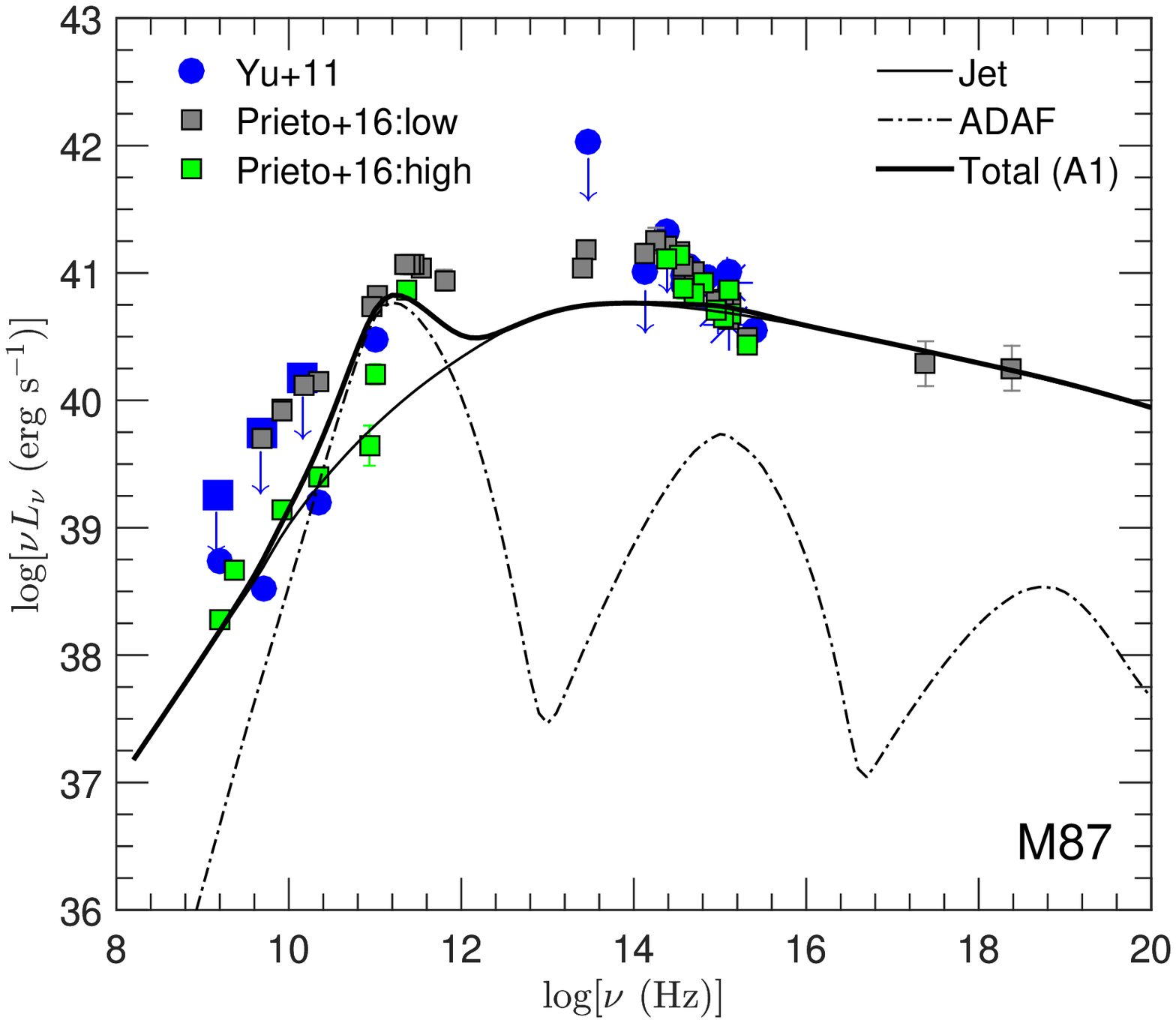}
\end{center}
\caption{\small The SED modeling of M87 by Model A1. The blue points (filled circles and squares) are taken from \citeauthor{Yu11} (\citeyear{Yu11}; see also \citealt{Maoz07}), the grey ones are from \citet{Prieto16}  with a low angular resolution of 0.4 arcsec (32 pc$\simeq5\times10^4\ R_{\rm S}$) aperture radius, while the the green ones are from the same paper with a high resolution of 0.15 arcsec. The dot-dashed line denotes the emission from the inner hot accretion flow,  the thin solid line denotes the jet component. The sum of the two components is denoted by the thick solid line.
}\label{fig:M87:sed}
\end{figure}

The SED modeling results for Model A1 are shown in Figure~\ref{fig:M87:sed}. The adopted fitting parameters are listed in Table~\ref{tab:para}. We can see that the mass accretion rate $\dot{m}_{\rm out}$ is well consistent with the Bondi accretion rate inferred from \emph{Chandra} observations. Other free parameters are also in their reasonable range.
The dot-dashed line in the figure denotes the emission from the ADAF, while the thin solid line is for the jet emission. The total radiation is denoted by  the thick solid line. In the model, the sub-mm radiation is dominated by the emission from the ADAF component, while the low-frequency radio and high-frequency optical-UV and X-ray radiation are dominated by the jet component.

\begin{figure}[htb]
\begin{center}
\includegraphics[width=0.45\textwidth]{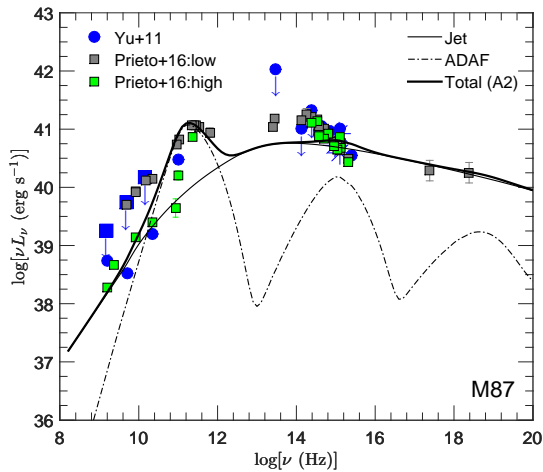}
\end{center}
\caption{\small Same as Figure~\ref{fig:M87:sed} but for Model A2 of M87.
}\label{fig:M87:sed2}
\end{figure}

The modeling results of Model A2 is shown in Figure~\ref{fig:M87:sed2}. The adopted parameters are listed in Table~\ref{tab:para}. An increase of the accretion rate at the Bondi radius by about a factor of 6 with a stronger outflow parameter $s$ can provide an equivalent fit to the data without any modification of the jet model parameters due to the same accretion rate in the inner region of the accretion flow. As in Model A1, the sub-mm emission is also dominated by the inner ADAF.
%The radial profile for the ADAF model are shown as blue dashed line in Figure~\ref{fig:M87:profile}.

\begin{figure}[htb]
\begin{center}
\includegraphics[width=0.45\textwidth]{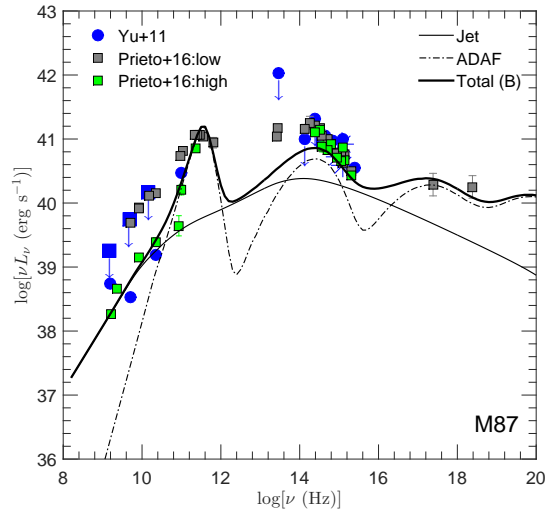}
\end{center}
\caption{\small Same as Figure~\ref{fig:M87:sed} but for Model B of M87. Different from Model A1 and Model A2, the X-ray emission in this model is  dominated by the inner ADAF.
}\label{fig:M87:sed3}
\end{figure}

In both of the two cases, the X-ray emission is dominated by the jet component. We therefore call Model A1 and A2 ``jet-dominated'' model. In the literature, there exists another type of model in the ADAF-jet framework, in which the X-ray emission is dominated by the ADAF. We call it ``ADAF-dominated'' model \citep{DiMatteo03,Nemmen14}. To calculate the corresponding RM of this type of model, we first need to reproduce the model. For this purpose, in Model B we adopt the similar parameters as in \citet{Nemmen14} since this model can be regarded as the ``updated version'' of \citet{DiMatteo03} model. The modeling results are shown in Figure~\ref{fig:M87:sed3} with the parameter values listed as model B in Table~\ref{tab:para}. We can see that the model can fit the SED satisfactorily. The sub-mm emission is again dominated by the ADAF component. Most of the parameters are the same with \citet{Nemmen14} except  the accretion rate $\dot{m}_{\rm out}$, which is larger in our model than that in \citet{Nemmen14} by  a factor of $\sim2$, and the fraction of shock energy transferring into electrons $\epsilon_{\rm e}$.  The main reason for the parameter difference is that we adopt a slightly higher X-ray luminosity in our SED data compared with that of \citet{Nemmen14}.

\begin{figure}[htb]
\begin{center}
\includegraphics[width=0.45\textwidth]{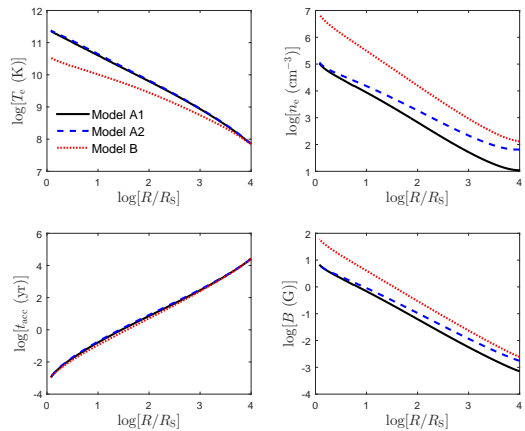}
\end{center}
\caption{\small The dynamical properties of ADAF for M87. The four panels present the radial profiles of the electron temperature $T_{\rm e}$, electron density $n_{\rm e}$, accretion timescale $t_{\rm acc}$ and magnetic field strength $B$, respectively. Different line styles correspond to different models.
}\label{fig:M87:profile}
\end{figure}

The radial profiles of the electron temperature $T_{\rm e}$, electron density $n_{\rm e}$, accretion time scale $t_{\rm acc}$, and magnetic field strength $B$ in the ADAF of Model A1, A2, and B are shown as the black solid, blue dashed, and red dotted lines in Figure~\ref{fig:M87:profile}. They will be used when we calculate the corresponding RM.

\subsection{SED Modeling of 3C 84}\label{sec:sed:3c84}

3C 84, one of brightest compact radio source, is associated with the nucleus of the giant elliptical galaxy NGC 1275 ($z = 0.0179$), which is Brightest Cluster Galaxy (BCG) of the prototypical ``cooling flow" cluster Abell 426 in Perseus \citep{Fabian94}. The mass of the central black hole is estimated to be $M_{\bullet}=8.0\times10^{8}\ M_{\odot}$ \citep{Scharwachter13} and the distance to us is $D=78.2\ {\rm Mpc}$\footnote{The redshift of 3C 84 is 0.0179 \citep{Strauss92} and the luminosity distance is derived from a $\Lambda$CDM cosmology with a Hubble constant $H_0 = 70\ {\rm km\ s^{-1} Mpc^{-1}}$ in a flat universe with $\Omega_{\rm m} = 0.29$.\\}. For the black hole mass and distance adopted here, $1\ {\rm mas}\simeq0.38\ {\rm pc}\simeq5.0\times10^{3}\ R_{\rm S}$.

The asymmetrical jets at both kpc \citep{Pedlar90} and pc scales \citep{Asada06} with the core-dominated morphology make it be classified as a Fanaroff-Riley type I (FR I) radio galaxy with the jet axis relatively aligning with our LOS.  The jets are mildly relativistic $(0.3c-0.5c)$ and are directed at an angle of $\theta\approx25^{\circ}\sim55^{\circ}$ by virtue of the inverted spectrum of the counter jet \citep{Vermeulen94},  the morphology and the brightness ratio of the northern and southern radio lobes \citep{Walker94}, the ratio of the apparent distances of northern and southern radio lobes \citep{Asada06}, and the broad-band spectral fitting \citep{Abdo09}. We adopt $\Gamma=2.0$ and $\theta=30^{\circ}$ in the SED modeling.

\begin{figure}[htb]
\begin{center}
\includegraphics[width=0.45\textwidth]{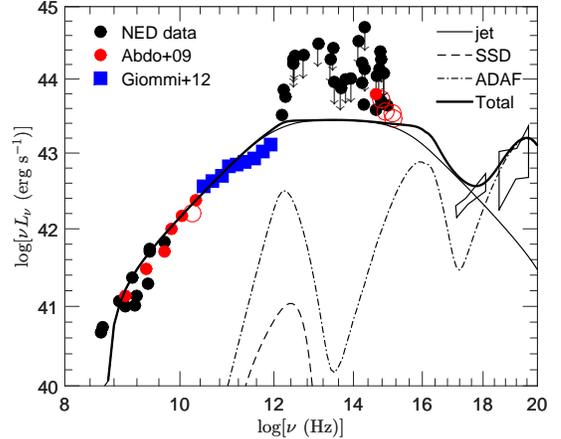}
\end{center}
\caption{\small The SED modeling of 3C 84 by Model A. The SED data are adopted from \citet{Abdo09} (red points), NED (black points and X-ray bow-ties), and \citet{Giommi12}(blue points). The IR-optical-UV bump is treated as an upper limit as other works \citep[e.g.,][]{Abdo09} since it may contain contamination from host galaxy.  The dot-dashed (dashed) line is the emission from the inner hot accretion flow (outer cold think disk), while the thin solid line corresponds to the jet component. The sum of the three components is presented as the thick solid line.
}\label{fig:3C84:sed}
\end{figure}

The multi-band SED data are mainly from \citet{Abdo09} and archival NASA/IPAC extragalactic database (NED)\footnote{http://ned.ipac.caltech.edu/}, as shown in Figure~\ref{fig:3C84:sed}. The bolometric luminosity is $4\times10^{44}\ {\rm erg\ s^{-1}}$, about $0.4\%$ of its Eddington luminosity $L_{\rm Edd}\approx1.0\times10^{47}\ {\rm erg\ s^{-1}}$. In combination with the apparent radio jet detection, it is thus reasonable to use the accretion-jet model to interpret its broad-band SED\footnote{The critical luminosity of an ADAF is roughly $L_{\rm bol,cr}=\epsilon\dot{M}_{\rm \bullet,cr}c^2\simeq0.03\%(\alpha/0.1)^2\dot{M}_{\rm Edd}c^2=2.7\%L_{\rm Edd}$ for the case of $\alpha=0.3$ and $\delta=0.5$ \citep{Xie12}. So the accretion mode can be in the ADAF regime.}. The IR-optical-UV bump is poorly constrained, which more likely contains host galaxy contamination. It is thus treated as an upper limit in our SED modeling.

As we have stated in \S2.1, numerical simulations of hot accretion flows show that the wind parameter $s$ is in the range of $0.3-1.0$. We consider two cases in this paper, {with $s=0.5$ in Model A} and $s=0.8$ in Model B. The other model parameters are listed in Table~\ref{tab:para}. The  modeling results of Model A are shown in Figure~\ref{fig:3C84:sed}. The results of Model B are similar thus are not shown here. But these two models give two different values of RM, as we will describe in \S3.2. The dot-dashed (dashed) line represents the emission from the ADAF (standard thin disk), while the thin solid line denotes the jet component. The total radiation is denoted by the thick solid line. As expected, the emission from the outer thin disk is not important for the overall SED due to the large truncation radius $r_{\rm tr}$. The {broad-band} emission is dominated by the jet except that the X-ray is attributed to the Comptonization of the thermal electrons in the ADAF.

%We note that the ADAF origin of the X-ray emission in 3C 84 is consistent with the \citet{YC05} prediction. According %to \citet{YC05}, for 3C 84, the critical X-ray luminosity below which the X-ray emission should be dominated by the %jet $L_{\rm X, crit}=1.4\times10^{-7}\ L_{\rm Edd}$. This is much lower than the observed 2-10 keV luminosity of 3C %84, which is $L_{\rm X}=2.1\times10^{-4}\ L_{\rm Edd}$ \citep{Verrecchia07}.

We note that 3C 84, with its X-ray luminosity $L_{\rm X}=2.1\times10^{-4}\ L_{\rm Edd}$ \citep{Verrecchia07} which is much higher than the critical X-ray luminosity $L_{\rm X, crit}=1.4\times10^{-7}\ L_{\rm Edd}$, is also
expected to have an ADAF origin for its X-ray emission, according to \citet{YC05}.

\begin{figure}[htb]
\begin{center}
\includegraphics[width=0.45\textwidth]{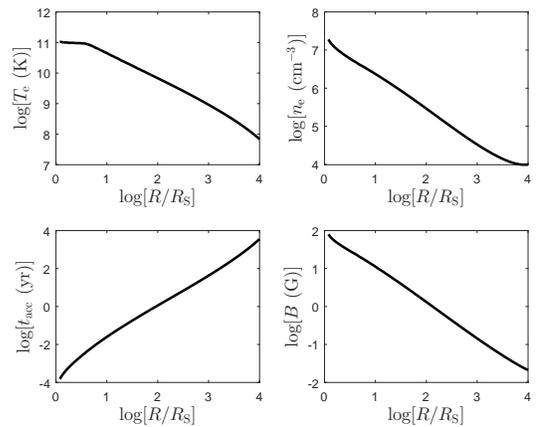}
\end{center}
\caption{\small The dynamical properties of ADAF of Model A of 3C 84. The four panels present the radial profiles of the electron temperature $T_{\rm e}$, electron density $n_{\rm e}$, accretion timescale $t_{\rm acc}$, and magnetic field strength $B$, respectively.
}\label{fig:3C84:profile}
\end{figure}

The dynamical profile of the electron temperature $T_{\rm e}$, electron density $n_{\rm e}$, accretion time scale $t_{\rm acc}$ and magnetic field strength $B$ of the ADAF of Model A are shown in four panels of Figure~\ref{fig:3C84:profile}.

\begin{table*}[htb]
\centering
\caption{Model parameters resulting from the SED fitting}
\label{tab:para}
\resizebox{0.85\textwidth}{!}{%
\begin{tabular}{ccccccccccccccc}
\hline\hline
    %\hline
  Object &	Model & $M_{\bullet}\ (M_{\odot})$ & $D$ (Mpc)	& $r_{\rm tr}\ (R_{\rm S})$ & $\dot{m}_{\rm out}(r_{\rm tr})^a$ & $s$ & $\dot{m}_{\rm jet}^a$ & $\Gamma$ & $\theta(^{\circ})$ & $\epsilon_{\rm e}$ & $\epsilon_{\rm B}$ & $p$ & Ref.\\
  (1) & (2)& (3) & (4) & (5) & (6) & (7) & (8) & (9) & (10) & (11) & (12) & (13) & (14)\\
  \hline
  \\
 %\hline
 %\\
     & A1 & $5.9\times10^9$& 16.4 & $-$ & $1.0\times10^{-4}$ & $0.3$ & $5.0\times10^{-9}$ & $6.0$ & $10$ & 0.001 & 0.008 & 2.35 & (1,2)\\
 M87 & A2 & $-$ & $-$ & $-$ & $6.0\times10^{-4}$ & $0.5$ & $5.0\times10^{-9}$ & $6.0$ & $10$ & 0.001 & 0.008 & 2.35 & $-$\\
     & B$^{b}$ & $-$ & $-$ & $-$ & $1.2\times10^{-3}$ & $0.1$ & $1.0\times10^{-8}$ & $6.0$ & $10$ & 0.0006 & 0.001 & 2.6 & $-$\\
 3C 84 & A & $8.0\times10^8$& $78.2$ & $10^4$ & $1.2\times10^{-2}$ & $0.5$ & $1.5\times10^{-5}$ & $2.0$ & $30$ & 0.10 & 0.05 & 2.8 & (3,4)\\
  %     & B & $-$& $-$ & $10^4$ & $1.2\times10^{-2}$ & $0.5$ & $9.0\times10^{-5}$ & $2.5$ & $50$ & 0.04 & 0.04 & 2.2 & $-$\\
        & B & $-$& $-$ & $10^4$ & $8.0\times10^{-2}$ & $0.8$ & $1.5\times10^{-5}$ & $2.0$ & $30$ & 0.10 & 0.05 & 2.8  & $-$\\
 %\\
  \hline\hline
\end{tabular}
}
\tablecomments{\small Column 1: name of the object. Column 2: model label. Column 3: mass of the BH. Column 4: distance of the object. Column 5: truncation radius in unit of $R_{\rm S}$. Column 6: dimensionless mass accretion rate at $r_{\rm tr}$. Column 7: outflow parameter. Column 8: mass loss rate in the jet. Column 9: Lorentz factor of the jet. Column 10: viewing angle of the jet. Column 11: fraction of shock energy entering into
electrons. Column 12: fraction of shock energy entering into the magnetic field. Column 13: spectral index of the power-law electrons. Column 14: references for the black hole mass and distance. \\
$^a$ The dimensionless mass accretion rate is defined as $\dot{m}=\dot{M}/\dot{M}_{\rm Edd}$, where $\dot{M}_{\rm Edd}\equiv22M_{\bullet}/(10^9M_{\odot})\ M_{\odot}\ {\rm yr^{-1}}$  is the Eddington accretion rate by assuming a radiative efficiency of 10 per cent.\\
$^b$ In addition, $\delta=0.01$, which is the fraction of the turbulent energy that directly heats the electrons as the AD model in \citet{Nemmen14}.\\
References: (1) \citet{Gebhardt09}; (2) \citet{Jordan05}; (3) \citet{Scharwachter13}; (4) \citet{Strauss92}.\\
}
\end{table*}

\section{Rotation Measure and The physical implications}

The fully corrected RM is given by \citep{Shcherbakov08,Huang11}

\begin{equation}\label{eq:rm}
  \textrm{RM} = 8.1\times10^5\int_{R_{0}}^{R_{\rm out}} g(X)\frac{K_{0}(\gamma_{\rm e}^{-1})}{K_{2}(\gamma_{\rm e}^{-1})}n_{\rm e}\textbf{\emph{B}}\cdot{\textbf{\emph{dl}}}~ \textrm{rad~m}^{-2}
\end{equation}
for the electron density $n_{\rm e}$ in units of cm$^{-3}$, the path length \textbf{\emph{dl}} in units of pc, the magnetic field \textbf{\emph{B}} in units of Gauss, and the dimensionless electron temperature $\gamma_{\rm e}$  defined as $\gamma_{\rm e}=kT_{\rm e}/m_{\rm e}c^2$.  The factor $J=\frac{K_{0}(\gamma_{\rm e}^{-1})}{K_{2}(\gamma_{\rm e}^{-1})}$ reduces to $J\rightarrow\log(\gamma_{\rm e})/2\gamma_{\rm e}^{2}$ as $\gamma_{\rm e}\rightarrow\infty$, and recovers to unity as $\gamma_{\rm e}\rightarrow0$.  This factor is due to the relativistic correction of electron mass \citep{Quataert00b}. Another factor $g(X)$ will suppress the RM in the low frequency band, which is not important here. For the high frequency approximation, $g(X)\simeq1$.

\subsection{Rotation Measure of M87}

Before calculating the RM contribution from the jet and the accretion flow, we first briefly discuss other possible contributions to the observed RM from  larger scales. \citeauthor{Algaba16} (\citeyear{Algaba16}; see also \citet{Owen90,Zavala03} for the RM measure at similar spatial scales) measured the RM at kpc scale using archival polarimetric VLA data at 8, 15, 22 and 43 GHz, and found that it is on the order of a few $\times10^2~ {\rm rad~m}^{-2}$. This basically present an order of magnitude estimation to the contribution to the RM by the interstellar medium and other large scale structure along our LOS. This RM may come from the wind launched from the accretion flow, since as we show below, the contribution of RM from the accretion flow is much higher.

%Even in the case of our LOS intersects with the foreground jet, its contribution should be minor, since such a small opening angle of the jet will lead to most of polarized emission being free of blocking out by the jet. Actually, the RM contribution from the jet is only ${\rm RM_{\rm jet,A1}} \approx3.2\times10^{3}\ {\rm rad~m}^{-2}$ for model A1 when  integrating Equation~\ref{eq:rm} along our LOS through the jet base to its surface.
%\begin{equation}\label{eq:rm:jet}
%  {\rm RM_{\rm jet,A}} \approx3.2\times10^{3}\ {\rm rad~m}^{-2},
%\end{equation}
%This value is much less than the observed $3\sigma$ upper limit value $7.5\times10^{5}\ {\rm rad~m}^{-2}$ \citep{Kuo14}.
%Because of a large fraction of the sub-mm emission coming from the inner ring of the jet, the point of the integration begins from the far side of the jet surface at the $10\ {R_{\rm S}}$ from the black hole to the near side of the jet (ref. to the LOS denoted as the black dashed line in Figure~\ref{fig:los}),  both of which are self-consistently determined by the jet geometry from our SED modeling. Since the observed RM lies in between $-7.5\times10^{5}$ and $3.4\times10^{5}\ {\rm rad~m}^{-2}$ \citep{Kuo14}, this small RM can actually be reconciled with the polarization observation.

\begin{figure*}[htb]\vspace{-1cm}
\begin{center}
\includegraphics[angle=90,width=0.85\textwidth]{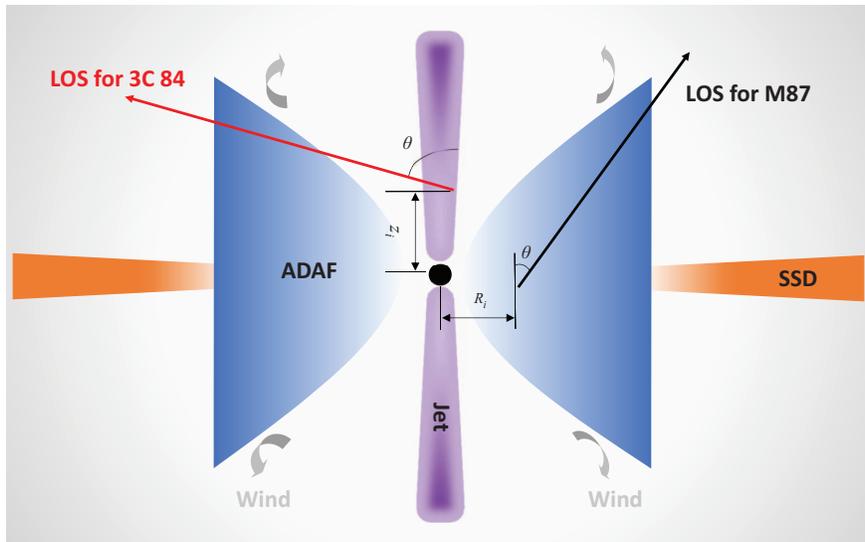}
\end{center}\vspace{-1.5cm}
\caption{\small The schematic figure for the geometry of the accretion-jet model and the calculation of RM. The model  consists of the outer thin disk (SSD), the inner hot accretion flow (ADAF), and the vertical
jet. The black solid line denotes the LOS for calculating the RM contribution from the ADAF of M87. For the case of 3C 84, since the sub-mm radiation mainly comes from the jet, the integration path is denoted as the red solid line which includes the contributions from  both the jet and ADAF.
}\label{fig:los}
\end{figure*}

Our SED modeling results suggest that the sub-mm emission, for which the RM is inferred from the polarization observation by \citet{Kuo14}, is mainly attributed to the ADAF component. Therefore, we focus on the RM contribution from the ADAF since the contribution of the jet is negligible\footnote{Since only a small fraction of accretion material in the inner region of the accretion flow is transferred to the jet, the plasma density in the jet is orders of magnitude lower than that in the ADAF and the associated magnetic field is also weaker. So the RM from the jet is insignificant compared with the ADAF. In addition, most of the sub-mm emission comes from the ADAF component, the jet contribution to the RM is thus negligible.}. To calculate the RM contribution from the accretion flow, we need information of the electron density, temperature and magnetic field distribution in the $r-\theta$ plane of the accretion flow. The electron density, temperature and magnetic field distribution along the $\theta$ direction are obtained by referring to the three dimensional MHD numerical simulations of accretion flows of \citet{DeVilliers05}. Fig. 3 in that paper shows that the density and gas
pressure decrease exponentially from the equatorial plane with polar angle.
%%The radial distribution of these quantities in the equatorial plane can be obtained from Figure~\ref{fig:M87:profile}.
The configuration of the magnetic field is somewhat uncertain. Although the MHD numerical simulation can give us some information, the result however in some degree depends on the initial configuration of the magnetic field. For simplicity, we assume a purely radial configuration in our calculation of RM. This would introduce an uncertainty of a factor of $\sim {\rm cos}(\alpha)$, with $\alpha$  being the inclination angle of the magnetic field line to the radial direction. MHD numerical simulations show that for our preferred inclination angle, the field reversal will further introduce a decrease of RM by a factor of $0.2\sim0.9$, depending on the initial configuration of the magnetic field in the simulation \citep{Sharma07}. Thus these two effects will not significantly change our result. In fact, in the {case of Sgr A$^\star$} the comparison of the calculation of RM between analytical calculation and MHD numerical simulation has shown a rough consistency  \citep{Sharma07}.

The polarized sub-mm radiation comes from different radius of the accretion flow. Therefore, when we calculate the RM, we should weight the RM from different radius with the emitted luminosity ($L$) emitted from the corresponding region. For simplicity, we use linear weighing here,
\begin{equation}\label{eq:rm:fluxdef}
  {\rm RM}=\frac{\sum_{R_{i}}{\rm RM}(R_{i})\times L(R_{i})}{\sum_{R_{i}}L(R_{i})},
\end{equation}
where $L(R_{i})$ is the luminosity of $i$th ring ($R_{i}\sim R_{i+1}$) in the accretion flow, and ${\rm RM}(R_{i})$ means the RM corresponding to the $i$th ring, $R_{i}$, of the accretion flow (refer to Figure~\ref{fig:los}).

Now we can calculate the RM from the ADAF  along our LOS based on Equations~(\ref{eq:rm}-\ref{eq:rm:fluxdef}). It gives %\footnote{Although the emitting plasma and the Faraday screen are both attributed to the accretion flow, we still treat it as external Faraday screen because the main contribution of the RM comes from a %large radius in the accretion flow, while the emission is dominated by the inner region.}
\begin{equation}\label{eq:rm:adaf}
  {\rm RM_{\rm A1}} \approx1.7\times10^{4}\ {\rm rad~m}^{-2},
\end{equation}
for the case of the inclination angle of $\theta=10^{\circ}$.
We have taken into account a series of emitting rings from the innermost region ($2~R_{\rm S}$) to the outer bounder ($R_{\rm Bondi}$) of the accretion flow to implement the luminosity-weighted RM calculation. We find that it is the inner ring that dominate the RM contribution since it dominates the sub-mm emission.
%The beginning integration point should be very close to the central black hole according to our SED modeling and VLBI observations \citep{Doeleman12,Akiyama15} (ref. to the LOS denoted as the black solid %line in Figure~\ref{fig:los}).
%Here we set $R_{0}=20\ R_{\rm S}$ and then test the dependence of the RM value on it below.
%The end point of the integration is assumed to be $R_{\rm out}=10^5\ {R_{\rm S}}$ and it is not important as the plasma is so tenuous and the magnetic field is extremely weak at  the outer boundary as shown in Figure~\ref{fig:M87:profile}.
For the case of \emph{internal} Faraday rotation considered here, namely, the emitting plasma and the Faraday screen are the same plasma, there is an additional correction factor of $\sim1/2$ \citep{Burn66,Cioffi80,Homan12}. But note that the RM contribution from the lower half plane of the accretion flow can add a correction factor of $\sim 2$ for the emitting sources from the corresponding regions. Therefore, such two modifications to the RM are canceled.

\begin{figure}[htb]
\begin{center}
\includegraphics[width=0.45\textwidth]{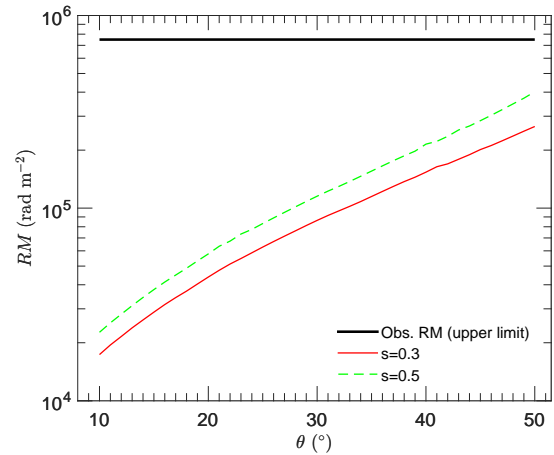}
\end{center}
\caption{\small Calculated RM from the ADAF as a function of the inclination angle for Model A1 and Model A2 of M87. The green dashed line shows the case of $s=0.3$ (Model A1) while the red line are for the case of $s=0.5$ (Model A2). The upper limit of the RM is shown as the thick black line.
}\label{fig:M87:RM}
\end{figure}

This RM value is close to the upper limit of the observed one. As there exists some room to adjust the inclination angle both from our SED modeling and observational constraints \citep{Biretta99,Acciari09,Wang09}, we have calculated the RM as a function of the viewing angle, which is shown as green line in Figure~\ref{fig:M87:RM}.
 %Note that we do not fit the SED data with different inclination angles $\theta$ again, which may have an impact on the spectra emitted from the jet, but only calculate the RM from the ADAF with different $\theta$. However, it should not be a serious problem since the RM from the jet is significantly smaller even though our LOS can intersect with the magnetized plasma in the jet.
It is easy to understand that a larger inclination angle leads to a larger RM since the LOS will intersect with more dense plasma in the accretion flow.
%The shaded region is simply because of the change of the beginning point of the integration $R_{0}$ from $20\ R_{\rm S}$ to $2\ R_{\rm S}$. We can see that the inner region of the accretion flow contributes insignificantly to the total RM since the RM from relativistic electrons in the inner region of the accretion flow is highly suppressed.
The observed upper limit of the RM is shown as the thick black solid line.
We can see that for the inclination angel range considered here, $\theta\lesssim50^{\circ}$,  it can be reconciled with the observed upper limit.
%This constraint is roughly consistent the observed range $\lesssim20^{\circ}$ for the inclination angle as mention above.
Therefore, the model A1 is well consistent with the RM observation.

We then calculate the RM for Model A2. The corresponding radial profiles of the ADAF are shown as blue dashed line in Figure~\ref{fig:M87:profile}. The result is
\begin{equation}
{\rm RM_{\rm A2}}\approx2.3\times10^{4}\ {\rm rad~m}^{-2}.
\end{equation}
The dependence of the RM on the inclination angle is shown as the red dashed line in Figure~\ref{fig:M87:RM}. We can see that Model A2 also passes the RM observational constraint.
%We find that for all the inclination angle with the consideration of the change of emission location in the accretion flow, the RM value is larger than the observed upper limit by a factor of about two. This solution is also consistent with the RM observation.
%It thus suggests that the strong outflow solution is disfavored by the polarization observations, if the sub-mm emission is dominated by the hot accretion flow.

Contrary to model A1 and A2, the X-ray emission in  model B is dominated by the accretion flow. Applying the radial profile obtained from the SED modeling in Section~\ref{sec:sed:87}, we can calculate the RM from the ADAF with $\theta=10^{\circ}$. The result is
\begin{equation}
 {\rm RM_{\rm B}}\approx1.3\times10^{8}\ {\rm rad~m}^{-2}.
\end{equation}
This value is larger than the observed value by two orders of magnitude. Given that  we roughly have ${\rm RM}\propto \dot{M}^{1.5}$, such a large value of RM is because of the large $\dot{M}$ at the Bondi radius and the small wind parameter $s$ compared with Model A1 and A2. They results in a much larger $\dot{M}$ at the inner region of the ADAF where sub-mm emission and also the RM are produced. Variation of $\theta$ can not significantly reduce the RM. Our assumption to the largely radial field and no field reversal should not significantly change the result, as we have argued in \S3.1. The model B is, therefore, strongly challenged by the RM observation.

%Interestingly,
%\citet{YC05} find that when the $2-10$ keV X-ray luminosity is lower than a critical value $L_{\rm X,crit}$, the X-ray emission of the system should originate from the jet rather than the ADAF. The %critical value is $L_{\rm X,crit}=1.0\times10^{-7}\ L_{\rm Edd}$ for M87, while its observed X-ray luminosity $L_{\rm X}=2.3\times10^{-8}\ L_{\rm Edd}$.
%The jet origin for the X-ray emission suggested by both our SED modeling and the RM constraint is well consistent with the predication of \citet{YC05}.

%This is the limiting case for which ADAF can dominate the sub-mm emission.

% A lower RM indicates that the jet component should contribute significantly to the sub-mm band.
\subsection{Rotation Measure of 3C 84}

The observed RM at sub-mm band in 3C 84 is $(8.7\pm6.9)\times10^{5}\ {\rm rad~m}^{-2}$. There are observational features which strongly suggest an \emph{external} origin of the RM, which means that the Faraday screen is not the same emitting plasma for the polarized emission. One is that the polarization angles in the sub-mm band seem to follow the $\lambda^2$ linearity quite well \citep{Plambeck14}. The second evidence is the variability. It is shown that the 1.3 mm flux density increased by a factor of about 1.6 from mid-2011 to mid-2013 \citep{Dutson14}. The polarization measurements in the same period, however, show no apparent systematic increase in RM \citep{Plambeck14}. It is impossible that the Faraday screen is associated with intracluster gas in the cluster Abell 426, since the typical RMs toward the cooling flow cluster are only in the range of $10^{3}\sim10^{4}\ {\rm rad~m}^{-2}$ \citep{Carilli02}. Specifically, the RM of 3C 84 at a distance 15 mas (5 pc) from the nucleus is $\simeq7000\ {\rm rad~m}^{-2}$\citep{Taylor06}, which is significantly lower than the observed RM. The RM contribution from the accretion wind should be much smaller than that from the accretion flow.  Therefore, in the following we will focus on the RM contributed by the jet and the accretion flow in our LOS.

From the SED modeling in Figure~\ref{fig:3C84:sed},  different from the case of M87, the sub-mm emission in 3C 84 is dominated by the jet component.
%In order to determine the location of emission in the jet, we decompose the jet emission into a series of rings with the heights from the black hole logarithmically uniform. The sub-mm emission from the %jet as a function of the height is shown in Figure~\ref{fig:3C84:jetrad}. The solid and dashed lines correspond to the emission at $\lambda=0.9\ {\rm mm}$ and $\lambda=1.3\ {\rm mm}$, respectively, which %show similar distribution along the jet. We can see that the sub-mm emission in the jet mainly comes from a relatively high latitude (${\rm a\ few}\times10^3\ {R_{\rm S}}$), compared with M87 for which it %is close to the jet base.
Then we can calculate the RM from the jet itself. Based on Equation~\ref{eq:rm} and taking into account the weighing of RM by the associated luminosity, we obtain ${\rm RM}=4.3\times10^4\ {\rm rad~m}^{-2}$. This is much lower than the observed value, $(8.7\pm6.9)\times10^{5}\ {\rm rad~m}^{-2}$.
%We consider different sites of the $0.9-1.3$ mm emission in the jet.
%By changing the altitude in the jet $z_{0}$ (or $R_{0}$) in Equation~\ref{eq:rm} from $100\ R_{\rm S}$ to $1500\ R_{\rm S}$ (ref. to the red solid line in Figure~\ref{fig:los} for the integrating path in %the jet), the RM is $(7.1\times10^4\sim3.2\times10^{2})\ {\rm rad~m}^{-2}$, which is significantly lower than the observed value, $(8.7\pm6.9)\times10^{5}\ {\rm rad~m}^{-2}$.
Of course this does not mean that the model is not correct, since our line of sight can pass through the accretion flow so we should take into account the contribution to RM from the ADAF.

We use the same method as described in \S3.1 to calculate the contribution to the RM by the ADAF. The radial profiles of the ADAF in Model A of 3C 84 have been shown in Figure~\ref{fig:3C84:profile}. Although the SED modeling sets the viewing angle to be $\theta=30^\circ$, we relax this constraint here to investigate the influence of $\theta$ on the RM from the ADAF, since we find that the constrain from the SED modeling on the value of $\theta$ is not strict\footnote{We find that the SED data can also be fitted with $\theta=50^{\circ}$, while the RM contribution from the jet does not change much, still far less than the observed value.}. The sub-mm emission comes from different part of the jet, so the corresponding LOS will pass through different part of the ADAF for a given viewing angle $\theta$. Therefore, same with the case of M87, we need to weight the contribution of RM by ADAF with the luminosity from different part of jet.  The luminosity-weighted RM is expressed as
\begin{equation}\label{eq:rm:fluxdef2}
  {\rm RM}=\frac{\sum_{z_{i}}{\rm RM}(z_{i})\times L(z_{i})}{\sum_{z_{i}}L(z_{i})},
\end{equation}
where $L(z_{i})$ is the luminosity emitted from the $i$th ring ($z_{i}\sim z_{i+1}$) in the jet, and ${\rm RM}(z_{i})$ is the contribution to the RM by  the accretion flow that the line of sight corresponding to $z(i)$ passes through.

\begin{figure}[htb]
\begin{center}
\includegraphics[width=0.45\textwidth]{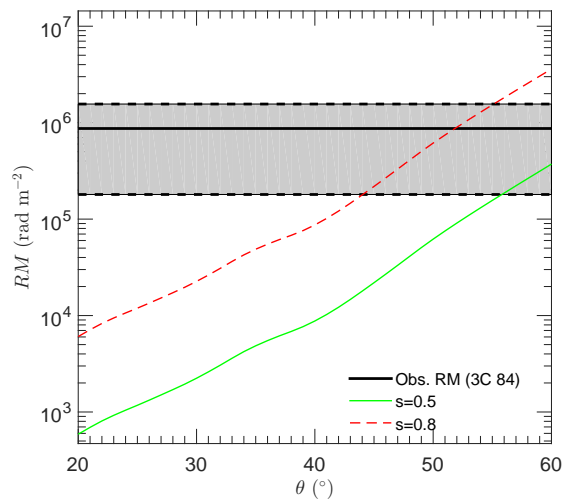}
\end{center}
\caption{\small RM from the ADAF as a function of the inclination angle for 3C 84. The observed RM with the $3\sigma$ limit is shown as the black line with the shaded area. The lines with different colors correspond to accretion-jet models with different outflow  parameters $s$.
}\label{fig:3C84:RM}
\end{figure}

The calculated RM for model A and B are shown in Figure~\ref{fig:3C84:RM}. The shaded region shows the observed RM value with the $3\sigma$ uncertainty. For the case of $s=0.5\sim0.8$, we can constrain the inclination angle to be $43^{\circ}\leq\theta\leq60^{\circ}$, which are roughly consistent with observational constraints mentioned above ($30^{\circ}\leq\theta\leq55^{\circ}$; \citealt{Vermeulen94,Walker94,Asada06}).

Different from the case of M87, for 3C 84, the polarized emission source and Faraday rotator are associated with different plasma, i.le., the jet and ADAF, respectively. Such an {\it external} origin of RM is well consistent with the observed correlation between the polarization angle and wavelength and variability feature of RM mentioned at the beginning of this section. The former is easy to understand. The stable RM value in the two-year timescale can be understood as follows.  As the RM is mainly contributed by the accretion flow, it is the accretion timescale that determines the variability timescale of RM. The accretion timescale $t_{\rm acc}=R/v_{\rm r}$, where $v_{\rm r}$ is the radial velocity, is shown as the bottom left panel in Figure~\ref{fig:3C84:profile}. It is on the order of $\sim{\rm 10\ yr}$ for $R\sim500\ R_{\rm S}$ where most of the sub-mm emission comes from. On the other hand, since the sub-mm emission comes from the jet, it is expected to be variable in shorter timescales.

\section{Summary}

The coupled ADAF-jet model is widely adopted in the literature to {model the LLAGNs} (see review by \citealt{Yuan14}). However, this model was mainly constructed by modeling {the multi-band SED} of the AGNs. As a consequence, there exist some model degeneracy, namely different models exist in this framework and they all can interpret the SED well. For example, for M87, two types of model exist in the literature in this framework, namely the ``ADAF-dominated'' model (e.g., \citealt{DiMatteo03,Nemmen14}) and {the ``jet-dominated'' model (e.g., \citealt{Yuan09,Yu11,Nemmen14})}. In both models, the sub-mm emission comes from the ADAF but the X-ray originates from the ADAF and jet, respectively.  The recent measurement of Faraday RM in M87 and 3C 84 supplies a powerful new constraint to theoretical models. The inferred RM is ${\rm RM}=(8.7\pm6.9)\times10^{5}\ {\rm rad~m}^{-2}$ (3$\sigma$ uncertainty; \citealt{Plambeck14}) for 3C 84 and  ${\rm RM}=(-2.1\pm5.4)\times10^{5}\ {\rm rad~m}^{-2}$ (3$\sigma$ uncertainty; \citealt{Kuo14}) for M87. In this paper, we first model the SED of these two sources and then calculate the predicted RM of the model and compare it with the observed value. The aims are to test the ADAF-jet models and break the model degeneracy. In our calculation of RM, we simply assume that the magnetic field is in radial direction. In reality, a field reversal or disordered field configuration almost must exist. So this is perhaps the largest uncertainty in our model. However, calculations based on numerical simulations show that these effects may result in a difference of a factor of few, thus not affect our conclusion \citep{Sharma07}.

For M87, in both the ``ADAF-dominated'' and ``jet-dominated'' models, the main contributor for the RM is the ADAF. While both of them can fit the SED well, we find that the predicted RM by the ``ADAF-dominated'' model is $\sim 10^8~ {\rm rad~m}^{-2}$, over two orders of magnitude higher than the observed value.  On the other hand, the predicted value of RM by the ``jet-dominated'' model is $\sim 10^4~{\rm rad~m}^{-2}$, well below the observed upper limit of RM. Thus the ``jet-dominated'' model passes the examination of the RM observation. The reason for the rather high RM predicted by the ``ADAF-dominated'' model is the high accretion rate in the model. The general relativistic effect cannot change the conclusion since most of the RM comes from the region a few$\times10~R_{\rm S}$.

For 3C 84, we first successfully interpret its SED using the ADAF-jet model. In our model, the sub-mm emission comes from the jet. But we find that the main contributor to RM is the ADAF. This is because our line of sight from the jet to the observer usually pass through the ADAF, while the density and magnetic field in the ADAF is much larger than those in the jet. Our detailed calculations show that the required inclination angle of the jet to explain the observed RM should be $43^{\circ}\leq\theta\leq60^{\circ}$ (ref. to Figure~\ref{fig:3C84:RM}). This is well consistent with other independent observational constraints. The ADAF-origin of the RM in our model can also explain two other important observational constraints of RM. One is that the polarization angle follows the $\lambda^2$ linearly quite well, which indicates an {\it external} origin of RM. The second is the variability timescales. Observations found that the RM is stable in the two-year timescale while the 1.3 mm emission increases significantly at the same period. This is because the accretion timescale of the ADAF, which determines the variability timescale of RM, is longer than 2 years; while we usually expect a fast variability for the emission from a jet.

\acknowledgments

We thank the referee for carefully reading and very useful comments on our manuscript.
This work was supported in part by the National Basic
Research Program of China (973 Program, grant 2014CB845800), the
Strategic Priority Research Program ``The Emergence of Cosmological
Structures¡± of the Chinese Academy of Sciences (grant XDB09000000)
and the Natural Science Foundation of China (grants 11133005 and 11573051).
FGX was also supported in part by the Youth Innovation Promotion Association of CAS (id. 2016243). This research has made use of the NASA/IPAC Extragalactic Database (NED)
which is operated by the Jet Propulsion Laboratory, California Institute of Technology,
under contract with the National Aeronautics and Space Administration.

\end{document}